\documentclass[preprint,showpacs,preprintnumbers,amsmath,amssymb]{revtex4}
\usepackage[dvips]{graphicx}
 
\begin{document}
\preprint{\normalsize UT-Komaba/10-4}
\title{A lattice study of $\mathcal{N}=2$ Landau-Ginzburg model \\ using a Nicolai map}
\author{Hiroki Kawai}
\email{hirokik@hep1.c.u-tokyp.ac.jp}
\author{Yoshio Kikukawa}
\email{kikukawa@hep1.c.u-tokyo.ac.jp}
\affiliation{Institute of Physics, the University of Tokyo, Tokyo 153-8902,  Japan}
\date{\today}

\begin{abstract}
It has been conjectured that the two-dimensional $\mathcal{N}=2$ Wess-Zumino model 
with a quasi-homogeneous superpotential provides the Landau-Ginzburg description of 
the $\mathcal{N}$=2 superconformal minimal models. For the cubic superpotential 
$W=\lambda\Phi^3/3$, it is expected that the Wess-Zumino model describes $A_{2}$ model 
 and the chiral superfield $\Phi$ shows the conformal weight $(h,\bar{h})=(1/6,1/6)$ 
at the IR fixed point. We study this conjecture by a lattice simulation, extracting the weight from 
the finite volume scaling of the susceptibility of the scalar component in $\Phi$. We adopt a 
lattice model with the overlap fermion, which possesses a Nicolai map and a discrete 
R-symmetry. We set $a\lambda=0.3$ and 
generate the scalar field configurations by solving the Nicolai map on $L \times L$ lattices 
in the range $L=18$ -- $32$. To solve the map, we use the Newton-Raphson algorithm 
with various initial configurations. 
The result is $1-h-\bar{h}=0.660\pm0.011$,  which is consistent with the conjecture within the statistical error, 
while a systematic error is estimated as less than 0.5 \%.  

\end{abstract}

\maketitle

\noindent
\textit{Introduction}---  
Two-dimensional $\mathcal{N}=2$ Wess-Zumino (WZ) model with a quasi-homogeneous superpotential is expected to describe, at the infrared (IR) fixed point,  
the $\mathcal{N}=2$ superconformal minimal models
\cite{Di Vecchia:1985iy,Boucher:1986bh,Cappelli:1986hf,Cappelli:1986ed, Gepner:1986hr,Gepner:1987qi}. 
This conjecture, which followed an analogous discussion for $\mathcal{N}=0$ by Zamolodochikov\cite{Zamo}, has been tested 
in various aspects
\cite{Kastor:1988ef, Howe:1989az, Vafa:1988uu, Lerche:1989uy,Cecotti:1989jc, Cecotti:1989gv, Cecotti:1990kz,Witten:1993jg}. 
If one could calculate correlation functions of the WZ model directly in the IR region, it would give us a further test of the conjecture. 
Since the coupling of the WZ model becomes strong 
in the IR region, this  certainly requires non-perturbative techniques.

%

Lattice methods can be useful for this purpose.
Although the realization of supersymmetry on the lattice is known 
to be difficult because of the lack of translation invariance  and 
the failure of the Leibniz rule\cite{Dondi:1976tx,Kato:2008sp}, 
one can preserve some part of the extended 
supersymmetries which are not directly related to the translation symmetry
\cite{Sakai:1983dg,Catterall:2001fr,Kikukawa:2002as,Catterall:2003uf,Giedt:2005ae,Bergner:2007pu}
\cite{Cohen:2003xe}
\cite{Sugino:2003yb} 
\cite{Catterall:2009it}. 
In such a lattice model,  in general,  extra fine tunnings are required to keep the model within the universality class of the target continuum theory. 
But, in some lower-dimensional models,  
the part of the supersymmetries preserved on the lattice, 
together with other lattice symmetries,  turn out to be sufficient
to suppress the extra relevant and marginal operators. 
The two-dimensional $\mathcal{N}=2$ WZ model is 
an example 
of such supersymmetric 
field theories that can be formulated successfully on the lattice in the above 
sense\cite{Sakai:1983dg,Catterall:2001fr,Kikukawa:2002as}, 
and 
can be studied non-perturbatively through lattice simulations. 

The purpose of this article is to provide a  non-perturbative numerical evidence 
for the above conjecture in the case of the simplest cubic superpotential $W(\Phi)=\lambda\Phi^3/3$ by simulating
the lattice WZ model which possesses the Nicolai map
\cite{Nicolai:1979nr,Parisi:1982ud, Cecotti:1982ad} 
and a discrete R-symmetry\cite{Kikukawa:2002as}. 
For the cubic superpotential, 
it is expected that the WZ model 
describes the $A_{2}$ model ($c=1$) and the chiral superfield $\Phi$ shows 
the conformal weight $(h,\bar{h})=(1/6,1/6)$ at the IR fixed point.  
We will extract the conformal weight of the chiral superfield $\Phi$ from the finite volume scaling 
of the susceptibility of the scalar component $\phi$ in $\Phi$. 
Interestingly, the $A_{2}$ model is also realized by the  Gaussian model ($c=1$) with the coupling 
constant $K$ at the $\mathcal{N}=2$ supersymmetric points,  $K= 1/12 \pi$ or $3/4\pi$ 
\cite{Waterson:1986ru,Friedan:1989yz}.
We will extract the coupling constant $K$ of the  Gaussian model by identifying
the phase factor of the scalar component   $\phi$ as the $2\pi$-periodic Gaussian field. 
This will provide a clear numerical evidence of the full recovery of $\mathcal{N}=2$ supersymmetry in the IR limit.


%

\vspace{.5em}
\noindent
\textit{Lattice formulation of WZ model}---  
We adopt the lattice WZ model which possesses a Nicolai map and a discrete R-symmetry\cite{Kikukawa:2002as}.
The model is formulated with overlap Dirac operator\cite{Neuberger:1997fp,Neuberger:1998wv}, 
$D= (1/a) \big[ 1+X / {\sqrt{X^{\dagger}X}} \big]$, where  
$X=1- (1/2)\big{[} (1+\gamma_{\mu})a\nabla_{\mu}^{+} + (1-\gamma_\mu) a\nabla_{\mu}^{-}\big{]}$
with the first forward, backward difference operators $\nabla_{\mu}^{\pm}$. 
$D$ satisfies 
the Ginsparg-Wilson relation\cite{Ginsparg:1981bj, Luscher:1998pqa}, 
$D\hat{\gamma_{3}}+\gamma_{3}D=0$ with $\hat{\gamma_{3}}=\gamma_{3}(1-aD)$, 
and it  can be expressed as
$D=\gamma_{1}S_{1}+\gamma_{2}S_{2}+ a T$
 in a spinor decomposition, 
defining three difference operators $S_1$, $S_2$ and $T$ in relation of 
$2T=-(S_1)^2-(S_2)^2+(aT)^2$. 
%
%
%
Then, the bosonic and fermionic actions are given by
\begin{eqnarray}
S_{\rm B}&=& a^2 \sum_{x}\big{\{}\phi^{\ast} (2T) \phi+W^{\ast\prime} (1-a^2T / 2)W^{\prime}\\
&&\qquad +W^{\prime}(-S_{1}+iS_{2})\phi 
      +W^{\ast\prime}(-S_{1}-iS_{2})\phi^{\ast}\big{\}}, \nonumber\\
S_{\rm F}&=& a^2 \sum_{x}\bar{\psi}\big{(}D+F(\phi)\big{)}\psi,
\end{eqnarray}
where $F(\phi)=\frac{1+\gamma_{3}}{2}W^{\prime\prime}\frac{1+\hat{\gamma_{3}}}{2}
+\frac{1-\gamma_{3}}{2}W^{\ast\prime\prime}\frac{1-\hat{\gamma_{3}}}{2}$
and $W'=\partial W(\phi)/\partial\phi$, $W''=\partial^2 W(\phi)/\partial\phi^2$. 
In the finite lattice of the volume $L \times L$, 
we adopt the periodic boundary condition for both bosonic and fermionic fields.
%

Throughout this article we consider the cubic superpotential $W(\Phi)=\lambda\Phi^{3}/3$. 
The coupling $\lambda$, which has the mass dimension one,  is the unique mass parameter of our model, besides the lattice spacing $a$. Thus $\lambda$ gives the scale under which the WZ model reduces to a conformal field theory. 
To see the conformal behavior on the lattice, we should prepare a large lattice size $ L/a  \gg (a\lambda)^{-1}$, 
while the continuum limit is $L/a \to \infty$ and $ a \lambda \to 0$. 

We summarize the  symmetries of our lattice model. 
First we note that the WZ model possesses the Nicolai map~\cite{Nicolai:1979nr}, 
which is explicitly given by
\begin{eqnarray}
\eta=W^{\prime}+a (\phi-\frac{a}{2}W^{\prime})T+(\phi^{\ast}-\frac{a}{2}W^{\ast\prime})(S_{1}+iS_{2}).\label{Nicolai map}
\end{eqnarray}
The Jacobian of this map from $\{ \phi, \phi^\ast\}$ to $\{ \eta, \eta^\ast\}$ precisely cancels
the overlap fermion determinant $|D+F(\phi)|$, while the bosonic action $S_{\rm B}$
is identical to the Gaussian weight $a^2 \sum_{x}|\eta(x)|^2$, which allows us to interpret 
$\eta$,  $\eta^\ast$ as the random white noises. 
 Thanks to this map, the lattice model has the following on-shell nilpotent supersymmetry 
 $Q$:
\begin{eqnarray}
&&Q\phi=-\bar{\psi}_{-},    \,\,\quad Q {\phi}^{\ast}=-\bar{\psi}_{+},\\
&& Q\psi_{+}=-\eta^{\ast},  \quad Q\psi_{-}=-\eta,\\
&& Q\bar{\psi}_{+}=0,         \,\,\qquad Q\bar{\psi}_{-}=0,
\end{eqnarray}
where we write $\psi=(\psi_1,\psi_2)^t$, $\bar{\psi}=(\bar{\psi}_{1},\bar{\psi}_{2})$, $\psi_{\pm}=(\psi_{1}\pm\psi_{2})/\sqrt{2}$ and $\bar{\psi}_{\pm}=(\bar{\psi}_{1}\mp\bar{\psi}_{2})/\sqrt{2}$. 
This lattice model also has a $Z_{3}$ R-symmetry. $S_{\rm F}$ has the $U(1)$ R-symmetry under 
$\phi\to e^{-2i\alpha}\phi$, $\psi\to e^{i\alpha\hat{\gamma_{3}}}\psi$, $\bar{\psi}\to\bar{\psi}e^{i\alpha\gamma_{3}}\label{chiral}$. But this symmetry is broken to $Z_{3}$ $(\alpha=n\pi/3, n\in\textbf{Z})$ by the last two terms in $S_{\rm B}$, which would be surface terms in the continuum limit.

With these symmetries, we can show  in the lattice perturbation theory that 
the desired continuum limit is achieved without extra fine-tunings. The redefined fields $\varphi\equiv\lambda\phi$, $\chi\equiv\lambda\psi$ are helpful for this discussion. In this notation, $\varphi$ has mass dimension 1 and $\chi$ has 3/2, and $\lambda$ is factorized as the overall factor $1/\lambda^2$ in the action. This overall factor counts the number of loops as the Planck constant does. Consider the generic radiative correction to an operator $\mathcal{O}$ of mass dimension $p\,(\geq0)$ in the action,
\begin{eqnarray}
\bigg{(}\frac{c_{0} a^{p-4}}{\lambda^2}+c_{1}a^{p-2}+c_{2}a^{p}+\cdots\bigg{)}\int {\mathrm d}^2 x  \, \mathcal{O}
\end{eqnarray}
where $c_{0}, c_{1},...$ are constants. The first, second and third terms represent the contributions at tree, one-loop and two-loop levels. In the continuum limit $a\to0$
, the corrections terminate at the two-loop level. At the tree level the lattice action agrees with that of the WZ model in the continuum limit. So we have to consider the operators with 
$p\leq2$. Such operators which  preserve the $Z_{3}$ R-symmetry and the fermion number are a constant and ${\varphi}^\ast\varphi$. But the constant has no effect on the path-integral and ${\varphi}^\ast\varphi$ is forbidden by the supersymmetry $Q$. Thus we do not need any extra fine-tunings to achieve the desired continuum limit, at least in the perturbation theory. 

We also note that by the $Z_{3}$ R-symmetry, 
the cubic quasi-homogeneous superpotential is uniquely singled out. 
The Yukawa coupling terms 
$\bar{\psi}(1+\gamma_{3})\phi^n(1+\hat{\gamma_{3}})\psi$ with 
$n\neq1$ and their conjugates, which may  appear in  the models other than the $A_{2}$ model, are not allowed by the $Z_{3}$ R-symmetry. 
 Therefore,  we do not  need to worry about operator mixings even at finite lattice spacing $a$.

Unfortunately, however,  $|D+F|$ can be negative. 
It is easily shown that $\gamma_{1}(D+F)\gamma_{1}=(D+F)^{\ast}$
in the basis $\gamma_{1}=\sigma_{3}$, $\gamma_{2}=-\sigma_{2}$, implying 
that every non-real eigenvalue of $D+F$ is doubly degenerated and $|D+F|$ is real. 
But the possibility of unpaired real negative eigenvalues can not be excluded. 
Then,  $|D+F| \, {\rm e}^{-S_{\rm B}}$ can not be interpreted as a 
positive probability weight. 
Therefore, 
our lattice WZ model generally faces the so-called sign problem 
in the standard simulation methods
like the hybrid Monte Carlo (HMC) method.

\vspace{.5em}
\noindent
\textit{Sampling configurations}---  Next we explain how we sample configrations. 
We utilize the Nicolai map~\cite{Beccaria:1998vi,Luscher:2009eq,Kadoh:2009sp}. 
Using the Nicolai map Eq.(\ref{Nicolai map}), 
one can rewrite the expectation value of  
the observable ${\cal O}$ 
in the following form, 
\begin{eqnarray}
\langle\mathcal{O}\rangle=\frac{\langle\sum_{i=1}^{N(\eta)}\mathcal{O}(\eta, \phi_{i})\, {\rm sgn}|D+F(\phi_{i})|\rangle_{\eta}}
{\langle\sum_{i=1}^{N(\eta)}{\rm sgn}|D+F(\phi_{i})|\rangle_{\eta}},\label{obs}
\end{eqnarray}
where $\langle\cdots\rangle_{\eta}$ denotes the average over the Gaussian white noise $\eta$.
$\{ \phi_i \}_{i=1,\cdots, N(\eta)}$
are the solutions of the Nicolai map Eq.(\ref{Nicolai map}) for a given $\eta$, 
which contribute to the observable (i.e.$|D+F(\phi_{i})|\neq0$), 
and $N(\eta)$ denotes the number of the solutions. 
These solutions exist discretely. To see this, suppose a solution $\phi+\delta\phi$ that is infinitesimally different from another solution $\phi$, $|D+F(\phi)|\neq0$. Then Eq.(\ref{Nicolai map}) means $0=(\text{Re}\delta\phi,i\text{Im}\delta\phi)(D+F(\phi))\Rightarrow\delta\phi=0$ and therefore the sigma symbol comes in Eq.(\ref{obs}). 

The expression Eq.(\ref{obs}) suggests us a possible interesting simulation method.  
Suppose one succeeds in solving the Nicolai map Eq.(\ref{Nicolai map}) for a given $\eta$
to obtain the sets $\{ \phi_i\}$ and $\{{\rm sgn}|D+F(\phi_{i})| \}$.  
Then, one can simulate the model by observing the numerator and the denominator 
separately, provided that signals for the denominator remain of order ${\cal O}(1)$.  
An advantage of this method is that the autocorrelation between samples completely disappears. 
One do not need the updating procedure like the molcular dynamics in the HMC algorithm.
Of course, one needs to evaluate $ {\rm sgn}|D+F(\phi_{i})| $, 
but it would not be demading in two dimensions for moderate lattice sizes $L/a$. 
%
To solve the Nicolai map numerically, one may
 use the Newton-Raphson algorithm with a globally convergent strategy~\cite{Numerical recipes}.
 A difficulty  is that one do not know $N(\eta)$ a priori.
What one may try then is to solve the Nicolai map with various initial configurations. 

%
We sample configurations with the above strategy in this work.  
For each noise $\eta$, we try the Newton-Raphson iterations from 100 initial $\phi$ configurations. 
Since almost all noises would 
be of order ${\cal O}(1)$, the solutions should be of order  
${\cal O}(1)$. So we generate these initial configurations by the standard normal distribution. 
And 
we suppose that the configurations obtained in this method exhaust solutions of the Nicolai map.

We check the quality of these samples by two tests. One is the Witten index~\cite{Witten:1982df}. 
Note that the denominator in Eq.(\ref{obs}),
$\Delta \equiv \langle \, \sum_{i=1}^{N(\eta)}{\rm sgn}|D+F(\phi_{i})| \, \rangle_{\eta}$, 
is the partition function correctly normalized 
into the Witten index~\cite{Cecotti:1981fu}.
For instance, in the massive free case $W(\Phi)=m\Phi^2/2$, the Nicolai map is just  
a linear equation with a coefficient matrix $D+m(1-aD/2)$. 
Since the determinant of this matrix is positive definite, the solution is unique and 
$\Delta$ indeed reproduces the known result in the continuum theory: $\langle+1\rangle_{\eta}=1$. 
In the present cubic case, it must be close to $2$ at least in the continuum limit. 
The other test is the Ward identity of the supersymmetry $Q$ of the lattice model. 
For the right hand side of Eq.(\ref{Nicolai map}), the on-shell supercharge acts as 
$Q\eta=\delta S/\delta\psi_{+}$, $Q\eta^{\ast}=\delta S/\delta\psi_{-}$.
Then $\langle Q(\cdots)\rangle=0$ and the Schwinger-Dyson equations imply the following Ward identities on the lattice:
\begin{eqnarray}
&& \langle\eta(x_1)\dotsb\eta(x_m)\eta^{\ast}(y_1)\dotsb\eta^{\ast}(y_n)\rangle\nonumber\\
&& \quad \qquad =\begin{cases}
0  &m\neq n\\
\sum_{\sigma}\Pi_{k=1}^{m}\delta_{x_k,y_{\sigma(k)}} \quad &m=n.
\end{cases}
\label{Ward}
\end{eqnarray}
For instance, the case $m,~n=1$ reads 
$\langle S_{\rm B}\rangle=L^2$.
%
An ideal case to pass these tests is 
when it happens that
$\sum_{i=1}^{N(\eta)}{\rm sgn}|D+F|=2$ over the noises. 
In this case, the Witten index $2$ is exactly reproduced.
In addition, 
from Eq.(\ref{obs}), $\langle\eta\dotsb\eta^{\ast}\dotsb\rangle$ reduces to $\langle\eta\dotsb\eta^{\ast}\dotsb\rangle_{\eta}$,   
and the latter reproduces the right hand side of Eq.(\ref{Ward}) because the noise is the standard normal distribution.

The sampling was carried out at $a\lambda=0.3$, on different $L\times L$ lattices with 
$L=18, 20, 22, 24, 26, 28, 30, 32$. We generated 320 noises on each $L$. 
The results, shown  in TABLE~\ref{Table 1.},  are classified into following 4 patterns. \textbf{A}: 2 solutions with positive fermion determinants. \textbf{B}: 4 solutions, one with negative and the rest with positive determinants. \textbf{C}: 1 solution with positive determinant. \textbf{D}: 3 solutions with positive determinants.
 Since almost all the noises belong to \textbf{A} or \textbf{B} where $\sum_{i=1}^{N(\eta)}{\rm sgn}|D+F|=2$, 
 the summation is almost 2 over the noises. 
 The Witten index $\Delta$ and the ratio $\delta=(\langle S_{B}\rangle-L^2)/L^2$ over these samples are also shown  
 in TABLE~\ref{Table 1.}.
From these results, 
one can estimate 
the systematic error in our simulation method 
as less than 0.5\%.  

%

\begin{table}
\caption{Samples}
\normalsize
\begin{center}
\label{Table 1.}
\begin{tabular}{c|cccccccc}\hline\hline
$L/a$ & 18 & 20 & 22 & 24 & 26 & 28 & 30 & 32 \\ \hline
\textbf{A} & 316 & 319 & 319 & 316 & 316 & 314 & 307 & 316 \\
\textbf{B} & 3 & 0 & 1 & 3 & 4 & 6 & 10 & 4 \\
\textbf{C} & 1 & 1 & 0 & 0 & 0 & 0 & 1 & 0 \\
\textbf{D} & 0 & 0 & 0 & 1 & 0 & 0 & 2 & 0 \\ \hline
$\Delta$ & 1.997 & 1.997 & 2 & 2.003 & 2 & 2 & 1.994 & 2 \\ \hline
$\delta$ [\%] & 0.3 & 0.0 & 0.1 & 0.4 & 0.4 & 0.4 & 0.4 & 0.2 \\ \hline \hline
\end{tabular}
\end{center}
\end{table}

\vspace{.5em}
\noindent
\textit{Numerical results}---  With these configrations of  the scalar component $\phi$ and 
the random source $\eta$, 
one can calculate the various correlation functions  in the WZ model. 
To extract the conformal weight of the chiral superfield $\Phi$, 
we observe the finite volume scaling of the susceptibility of the scalar component $\phi$, 
\begin{equation}
\chi_{\phi} \, \equiv   \, a^2 \sum_{|x|\geq3a}\langle\phi(x)\phi^{\ast}(0)\rangle.
\end{equation}
Here we omit the contribution from  the correlations at the distances shorter 
than $\lambda^{-1}=a/0.3\simeq3a$, as an improvement of the observable. 
%
If 
the chiral superfield $\Phi$ really shows the scaling behavior with the conformal weight $(h,\bar{h})$, 
and 
the correlation function $\langle\phi(x)\phi^{\ast}(0)\rangle$ scales  in the IR region, 
the finite volume scaling of $\chi_{\phi}$ in the continuum limit should read
\begin{eqnarray}
\chi_{\phi} \, \, \,   \propto \, \, \,  \int_{V} {\mathrm d}^2x \frac{1}{|x|^{2h+2\bar{h}}} \, \, \,  \propto  \, \, \, 
V^{1-h-\bar{h}},
\end{eqnarray}
where $V=L^2$ is the system volume. Using this relation, we read $1-h-\bar{h}$ from the slope of the $\text{ln}\chi_{\phi}$-$\text{ln}L^2$ plots. 
%

\begin{figure}[t]
\centering
\includegraphics[width=5.1cm,angle=-90]{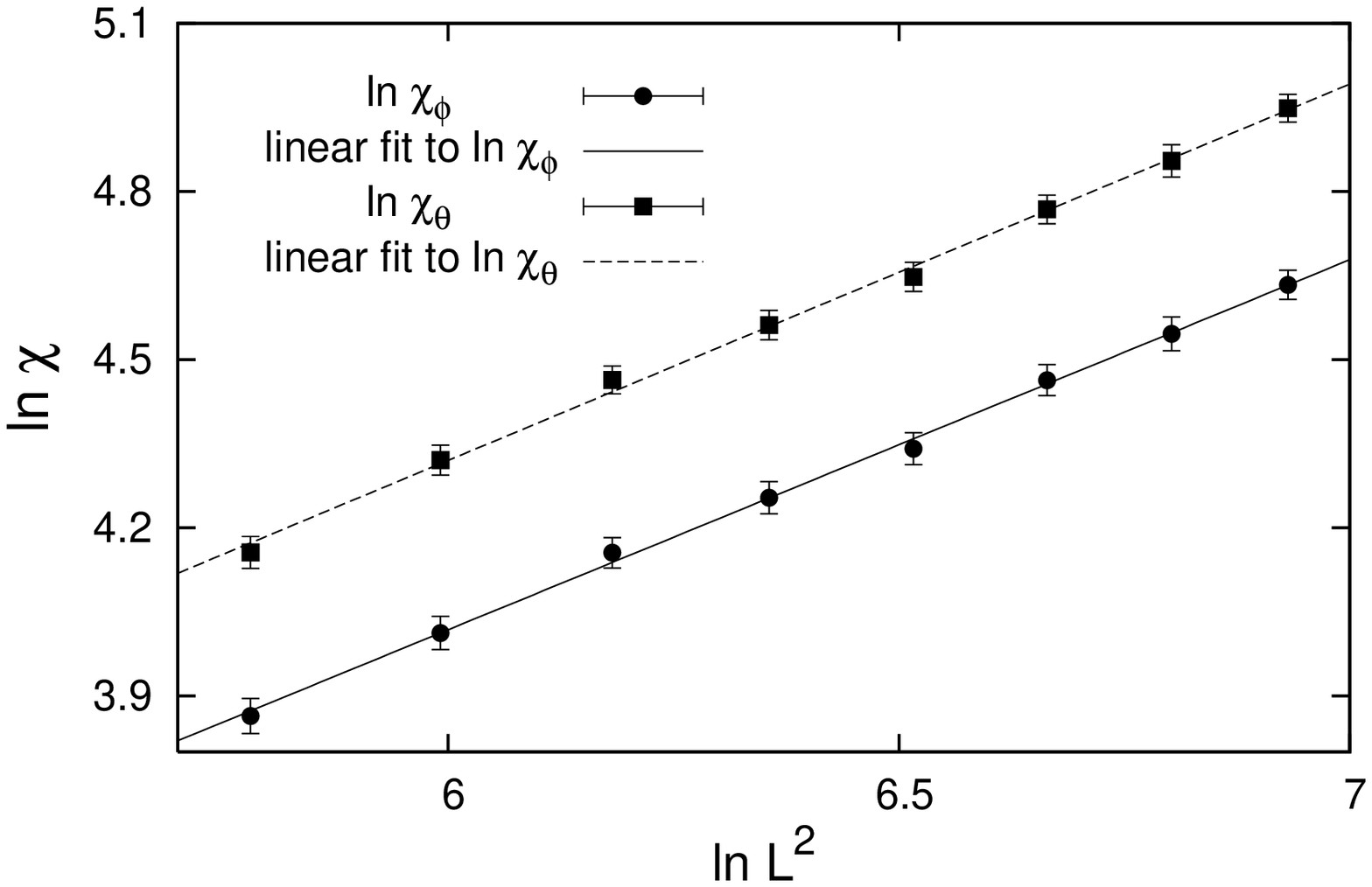}
\caption{$\text{ln}\chi$ - $\text{ln}L^2$ plots}
\label{fig.1}
\end{figure}

The numerical results  are shown in FIG.~\ref{fig.1}. The error bars show the statistical errors.
The solid  line in FIG.~\ref{fig.1} is a fit to the  plots by   least-square-method
and 
the slope gives $1-h-\bar{h}=0.660\pm0.011$. 
Our result is consistent with the value expected by the conjecture, $1-h-\bar{h}=2/3=0.666\ldots$.
This is our main result.

%


%

%


For a further support of the conjecture, we will also extract the coupling constant $K$ of the  Gaussian model as follows.
In the continuum limit, the WZ model has $U$(1) R-symmetry. 
Although this chiral symmetry is not  broken spontaneously, 
according to Coleman's theorem\cite{Coleman},  there may appear 
massive fermions and bosons in the spectrum and may decouple in the IR limit,  
leaving only massless degrees of freedom\cite{Witten78}. 
If  one writes $\phi=|\phi| {\rm e}^{i\theta}$, the R-symmetry is given by $\theta\to\theta-2\alpha$.
Then, the modulus $|\phi|$ and the new fermion $\chi= {\rm e}^{i \gamma_3 \theta/2} \, \psi $, 
$\bar \chi = \bar \psi \, {\rm e}^{i \gamma_3 \theta/2}$ are all singlets
and may aquire masses.  These degrees of freedom may decouple in the IR limit, leaving $\theta$ as low energy degrees of freedom. 
The IR effective action of $\theta$ may be given by
\begin{eqnarray}
S_{\rm eff}=\frac{K}{2}\int {\mathrm d}^2 x (\nabla\theta)^2 \qquad (\theta\sim\theta+2\pi\label{eff}), 
\end{eqnarray}
where $K$ is an effective coupling constant. 
Other chiral symmetric terms are irrelevant.


The value of $K$ will provide us a criterion for the full recovery of $\mathcal{N}=2$ 
superconformal symmetry in the IR limit.
The Gaussian model can reproduce $\mathcal{N}$=2 superconformal algebra, because
both bosonic and fermionic components of the chiral superfield can be constructed from 
the single bosonic field $\theta$, and the model contains indeed 
the superconformal stress-energy tensor  at the two values 
$K=1/12\pi$ and $3/4\pi$ \cite{Waterson:1986ru,Friedan:1989yz}. 
U(1) R-charges of $\phi$ and $\psi$ in the WZ model suggest that $K=3/4\pi$
is realized in our case. 

%

%

To extract the value of the coupling constant $K$, 
we again plot the finite volume scaling of the susceptibility 
\begin{equation}
\chi_{\theta} \, \equiv  \,   a^2 \sum_{|x|\geq3a}\langle e^{i\theta(x)}e^{-i\theta(0)}\rangle.
\end{equation}
 If the above scenario works, in the IR region, the operator ${\rm e}^{i\theta}$ becomes the vertex operator with the weight $(h,\bar{h})=(1/8\pi K,1/8\pi K)$. Then, 
one expects $\chi_{\theta}\propto V^{0.666\ldots}$. The scaling dimension is identical with $\chi_\phi$. 

The numerical results are shown also in FIG.~\ref{fig.1}. 
The solid line is again a fit to the plots. The slope is $0.671\pm0.014$ and therefore $K=0.242\pm0.010$, which is consistent with $K=3/4\pi=0.238\ldots$. Thus, our numerical results support the conjecture, 
 including the fact $c=1$. 
In addition, since $K=3/4\pi$ is the $\mathcal{N}=2$ supersymmetric point of the Gaussian model, 
this result implies the restoration of all supersymmetries 
in the IR limit (at long distances) 
without fine tunning.


\vspace{.5em}
\noindent
\textit{Acknowledgments}.---  
The authors would like to thank D.~Kadoh, M.~Kato, S.~Komatsu, J.~Noguchi, F.~Sugino, H.~Suzuki and T.~Yoneya for helpful discussions. 
The numerical calculations were carried out by T2K Open Super Computer system of the University of Tokyo. 
Y.K. is supported in part by Grant-in-Aid for Scientific Research No.~21540258, ~21105503.

 \end{document}